\documentstyle[12pt,psfig]{report}
\begin{document}

\thispagestyle{empty}

\begin{center}
\begin{center}
{ \LARGE \bf Multicolor photometry of the GRB970508 optical remnant}
\end{center}
\vspace{1.cm}

{ \Large \bf   V.V. Sokolov, A.I. Kopylov, S.V. Zharikov \\

{\large Special Astrophysical Observatory of RAS,
Karachai-Cherkessia, Nizhnij Arkhyz, 357147 Russia; sokolov,akop,zhar@sao.ru }
\vspace {0.25cm}

E. Costa$^{1}$, M. Feroci$^{1}$, 
L. Nicastro$^{2}$, E. Palazzi$^{2}$ }   \\

{\large $^1$ Istituto di Astrofisica Spaziale CNR, 00044 Frascati, Italy \\
$^2$ Istituto Tecnologie e Studio Radiazioni Extraterrestri CNR, 40129 Bologna, 
Italy \\ }
\end{center}
\vspace{1cm}

{\Large
\bf
We report results of follow-up multicolor photometry
of the optical variable source that is a probable remnant
of the gamma-ray burst GRB970508 discovered by the BeppoSAX satellite
(IAUC 6649).
Observations were carried out in Johnson-Kron-Cousins $\bf BVR_{c}I_{c}$ system
with the 1-m and 6-m telescopes of SAO RAS.
Between the 2nd and the 5th day after the burst a fading of the remnant is
well fitted with an exponential law in all four bands.
During this period the `broadband spectrum' of the object was unchanged and
can be approximated by a power-law,
 \fbox{ $\bf  F_{\nu}\propto\nu^{ -1.1} $}.
After the 5th day the decline of brightness is slowed down.
In the $\bf R_c$ band until the 
32nd day, 
the light curve can be described
by a power-law relation, \fbox{ $\bf F_t \propto t^{ -1.2}$}.
}

\clearpage
\begin{center}
{ \underline{\LARGE \bf THE SEARCH FOR AN OPTICAL}
 \underline{\LARGE \bf COUNTERPART}}
\end{center}
\large \bf

  The first coordinates of $10'$ radius error box 
for the GRB970508 (May 8.904 UT) were received 
in SAO RAS from the BeppoSAX team (by phone) on  May 9.05 UT.
At that time observations were not possible because of beginning of 
morning twilight.
On May 9.3 UT the refined coordinates of $5'$ radius error box were received 
by e-mail from the BeppoSAX team. 

The search for optical counterpart began with the
1-m telescope on May 9.74 UT.
The 5' error box for the  GRB970508 localization was
completely covered with the CCD mosaic of 29 images
in $\bf R_c$ band with 300 and 600 sec exposure times.
The CCD photometer at the 1-m (Zeiss-1000) telescope is equipped with
a ISD015A chip
of $\bf 520\times580$ pixels
corresponding to the field of view $\bf 2.0^{\prime}\times3.^{\prime}6$.
The images  from the 1-m telescope were compared to the corresponding fields
of the Digitized Sky Servey (DSS).
 No new bright object was found up to the DSS limit for this field.

On the next night, May 10/11 a better position was available:
$\bf \alpha_{2000}=06^h53^m28^s $; $\bf \delta_{2000}=+79^o17^{\prime}.4$ with
a $3'$ error radius (99\% confidence level).
Photometric observations of GRB970508 field were then continued with the 6-m
telescope with a CCD photometer
installed at the Primary Focus.
The CCD chip ``Electron ISD017A'' was used; its format of $\bf 1040\times1160$ pixels
corresponds to the field of view of
$\bf 2.^{\prime}38\times2.^{\prime}66$.
A $\bf 2\times2$ binning mode was employed,
so that each of the
$\bf 520\times580$ zoomed pixels (referred to as `pixels' hereafter)
has angular size of
$\bf 0.^{\prime\prime}274 \times 0.^{\prime\prime}274$.
The gain is  $\bf 2.3 e^-$ per DN (Data Number).
The readout noise is about $\bf 10e^-$.

The first image on the 6-m telescope was obtained on May 10.76 UT  and
a variable object was discovered using our data from the 1-m telescope.
Its brightness  from May 9.85 UT to May 10.76 UT increased
about 1.5 magnitudes. This object was first reported by H. Bond as a possible
optical counterpart of GRB970508 (IAUC6654) but was independently
found in our data only about 0.5 day later.
Log of observations of GRB970508 remnant in SAO RAS during the first 5 days
after the burst are given in Table 1.
Total exposures in seconds are given.

\vspace{1.5cm}
\begin{table}[h]
\caption[]{\large \bf Log of observations of GRB970508 remnant in  May.}
\begin{center}
\begin{tabular}{ccccccc}
\hline
night, May &  day,  UT   & telescope  &  B  &   V     &  $\bf R_c$      &    $\bf I_c$    \\  \hline
09/10 &  09.75 & 1-m  &     &         &  300    &         \\
 -"-  &  09.85 & -"-  &     &         &  600    &         \\  \hline
10/11 &  10.77 & 6-m  & 300 &    200  &  100    &  300    \\
 -"-  &  10.93 & -"-  & 300 &    200  &  100    &  300    \\
11/12 &  11.76 & -"-  & 450 &    300  &  150    &  450    \\
12/13 &  12.87 & -"-  & 450 &   600   &  150    & 900   \\
13/14 &  13.88 & -"-  &1200 &   600   & 450     &  450    \\  \hline
\end{tabular}
\end{center}
\end{table}

\clearpage
\begin{center}
{ \underline{\LARGE \bf  PHOTOMETRY}}
\end{center}

Observations were carried out with filters closely matching
the $\bf BVR_{c}I_{c}$ Johnson-Kron-Cousins system.

The data were processed using the ESO-MIDAS software.
Standard data reduction includes subtraction of the bias,
flat-fielding and removing of cosmic particle traces.

Photometric conditions
remained stable during two nights of May 13/14  and May 21/22.

\underline{Four bright stars (Fig. 2)} in the GRB970508 field were used as
secondary photometric standards.
Magnitudes of these stars
were determined on May 13/14 night with good photometric
conditions using four standard stars in the field of
PG1657+078 (Landolt, 1992).
Zero-point errors are better than $\bf 0.05^m$.
Coordinates and magnitudes of secondary photometric standards
are  given in Table 3. Our $\bf R_c$ magnitudes of
stars 2, 3, 4 are $\bf 0.20\pm0.01$ higher than the magnitudes measured
by Schaefer et al. (1997).

\clearpage
\begin{center}
{ \underline{\LARGE \bf RESULTS}}
\end{center}

Johnson-Kron-Cousins magnitudes
with its errors for GRB970508 optical counterpart are given
in Table 2.

{\LARGE \bf 1)} In the period of May   $\sim$9.13 UT (Castro-Tirado et al.,
 1997;  Djorgovski et al., 1997) to May 9.85 UT
the $\bf R_c$ brightness  of the object seems to remain constant.
\vspace{0.25cm}

{\LARGE \bf 2)} Object brightness in $\bf R_c$ band from May 9.85 UT to
 May 10.76 UT increased 1.5 magnitudes.
The magnitude increase rate using ours 1-m data and the data from Palomar
(Djorgovski et al., 1997)
 amount to 0.12 magnitude per hour.
\vspace{0.25cm}

{\LARGE \bf 3)} The brightness maximum was  $\bf t_{max}\approx1.5$ day
 after the burst.
 On May 10.76 UT the $\bf R_c$ magnitude was 19.70 and
since $\bf \approx$ May 10.76 UT a decline of brightness began.
\vspace{0.25cm}

{\LARGE \bf 4)} Measurements of the `broadband spectrum' on this stage
(2-5 days) of fading correspond to an exponential law in all bands:

\bigskip
\begin{center}
\fbox{\begin{minipage}{15cm}\begin{center}
  { $\bf  B = 19.689(\pm0.036)+0.452(\pm0.014)(t-t_0) $} \\
  { $\bf  V = 19.264(\pm0.053)+0.449(\pm0.020)(t-t_0) $} \\
  { $\bf  R_c = 18.874(\pm0.029)+0.443(\pm0.011)(t-t_0) $} \\
  { $\bf  I_c = 18.355(\pm0.050)+0.450(\pm0.019)(t-t_0) $}
\end{center} \end{minipage}
}
\end{center}
\bigskip
where  $\bf ( t-t_0$) is in days.

 The light curve of optical counterpart during the first 5 days after the burst
is shown in Figure 4.

The spectrum of the object was close to the power-law and its slope
$\bf  F_{\nu}\propto\nu^{ -1.2} $ did not change in time.
      $$\bf  (B-V)=0.43, \ \ (V-R_c)=0.39, \ \ (R_c-I_c)=0.52 $$

The account for the galactic absorption $\bf  E(B-V)=0.03  $ gives
\fbox{ $\bf  F_{\nu}\propto\nu^{ -1.1} $}  and the following
color indeces:
       $$\bf  (B-V)_0=0.40, \ \ (V-R_c)_0=0.37, \ \ (R_c-I_c)_0=0.50 $$
{\LARGE \bf 5)} The observations of the object on
May 22.00 UT I Jun 09.60 UT have shown  that
after 5 days the exponential law of brightness fading is changed to
a power-law \fbox{ $\bf F_{t}\propto t^{ -1.2}$}.
Figure 5 shows the light curve.


\bigskip
\bigskip
\begin{center}
{ \underline{\LARGE \bf CONCLUSIONS}}
\end{center}

The data obtained with the 1-m and 6-m telescopes
SAO RAS allows to divide the brightness change curve into three
stages:

\begin{enumerate}
\item the increase of brightness on the scale of about one day;

\item the exponential brightness fall during about 4 days with
the conservation of broadband power-law spectrum;

\item the further slowing down of the brightness fading
according to a power-law.
\end{enumerate}
\clearpage
\begin{center}
{\Large \bf Results of photometry of GRB970508 optical remnant in SAO RAS}
\end{center}
\vspace{-1.cm}

{\normalsize \bf
\begin{table*}[h]

\begin{center}
\begin{tabular}{lrccccccccccccc}
\hline
  UT & $\bf t-t_0$     &       B   & $\bf \sigma B$ &   V   & $\bf \sigma V$    &   $\bf R_c$   &$\bf \sigma R_c$    &      $\bf I_c$    & $\bf \sigma I_c$
  &     $\bf B-V$  &     $\bf V-R_c$  &    $\bf R_c-I_c$  \\ \hline
 May   &              &     &       &     &        &     &           &    &           &          &         \\
 9.745 & 0.841&       &     &       &     &  21.19 &0.25 &           &     &          &          &         \\
 9.848 & 0.944&       &     &       &     &  21.13 &0.18 &           &     &          &          &         \\
 10.77 & 1.866& 20.50 & 0.03& 20.06 &0.03 &  19.70 &0.03 &    19.19  &0.04 &     0.44 &     0.36 &    0.52 \\
 10.93 & 2.026& 20.60 & 0.03& 20.22 &0.03 &  19.80 &0.03 &    19.30  &0.03 &     0.38 &     0.42 &    0.50 \\
 11.76 & 2.856& 21.03 & 0.04& 20.52 &0.03 &  20.10 &0.03 &    19.58  &0.04 &     0.51 &     0.43 &    0.52 \\
 12.87 & 3.966& 21.48 & 0.06& 21.10 &0.04 &  20.63 &0.05 &    20.19  &0.06 &     0.38 &     0.47 &    0.45 \\
 13.88 & 4.976& 21.92 & 0.07& 21.47 &0.05 &  21.09 &0.07 &    20.58  &0.09 &     0.45 &     0.38 &    0.51 \\
 22.00 &13.096&       &     &       &     &  22.20 &0.15 &           &     &          &          &         \\
 Jun.  &              &     &       &     &              &           &     &          &          &         \\
 9.60  &31.696&       &     &       &     &  23.28 &0.10 &           &     &          &          &         \\ \hline
\end{tabular}
\end{center}
\label{mag}
\end{table*}

\begin{center}
{ \underline{\large \bf REFERENCES}}
\end{center}
\vspace{-0.5cm}

\parindent=0pt
Bond H.E., IAU Circ No. 6654 (1997). \\
Castro-Tirado A. J. et al.,   IAU Circ No. 6657 (1997). \\
Costa E. et al., IAU Circ No. 6649 (1997). \\
Djorgovski S. et al., Nature, Vol. 387, 876 (1997). \\
Landolt A. U., Astron. J., 104, 340 (1992). \\
Metzger M. R. et al., IAU Circ No. 6676 (1997). \\
Mignoli M et al.,  IAU Circ No. 6661 (1997). \\
Fruchter A. et al.,  IAU Circ No. 6674 (1997). \\
Schaefer B et al.,   IAU Circ No. 6658 (1997).
}
\clearpage
\begin{figure*}[t]
\centerline{
   \vbox{\psfig{figure=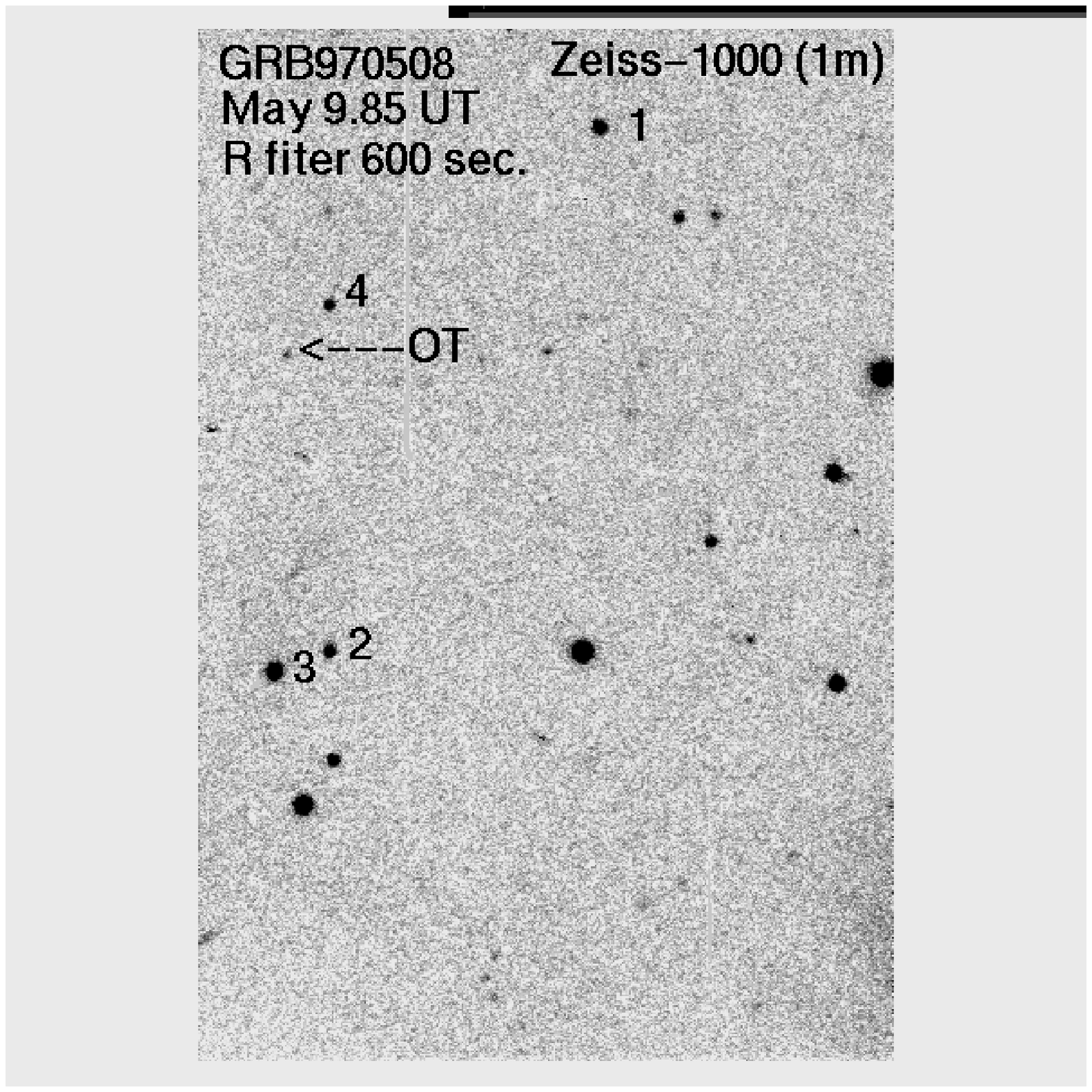,width=15cm,%
 bbllx=145pt,bblly=310pt,bburx=440pt,bbury=745pt,clip=}}
}
\caption{\bf Field  of GRB970508 optical counterpart from the 1-m telescope (Zeiss-1000).
}
\label{Zeiss}
\end{figure*}

\clearpage
\begin{figure*}[t]
\centerline{
   \vbox{\psfig{figure=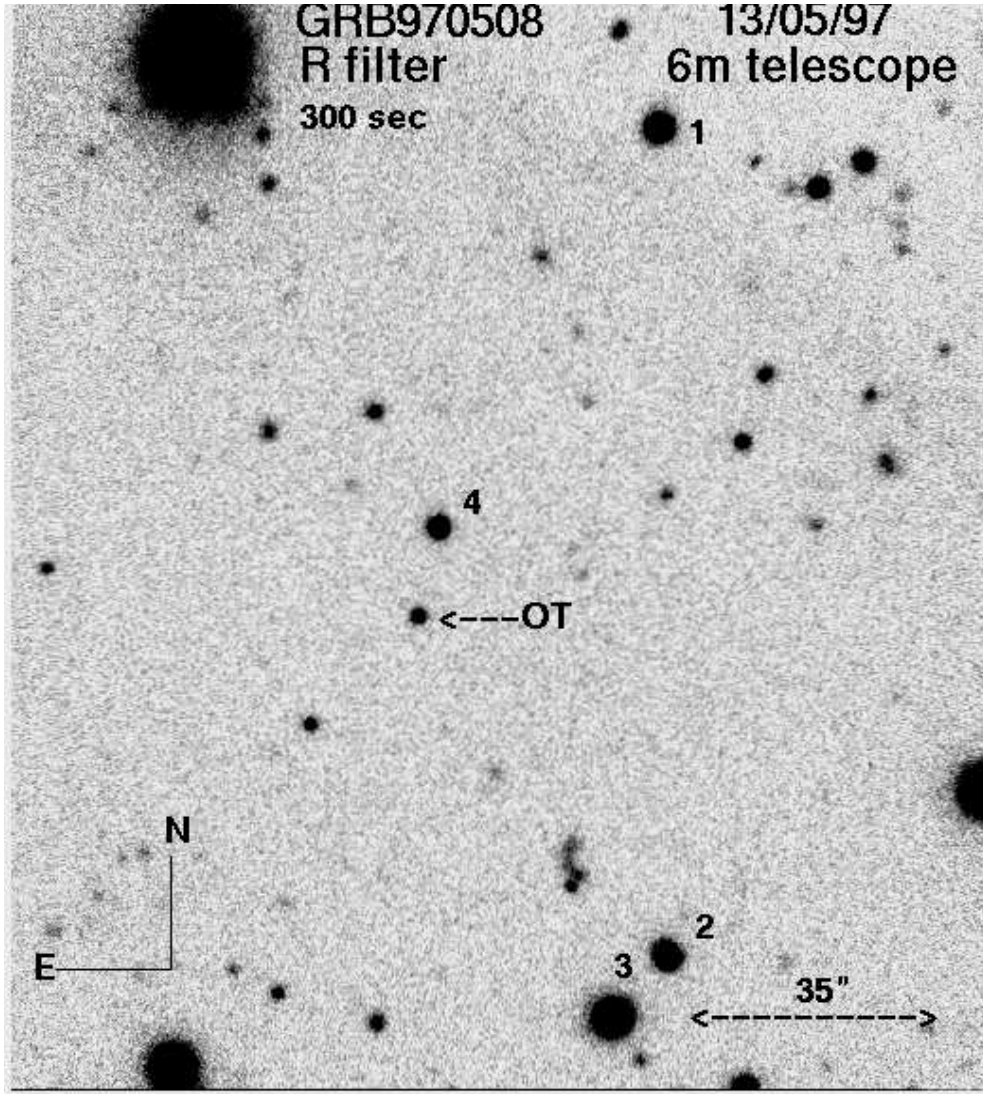,width=13cm,%
 bbllx=95pt,bblly=305pt,bburx=485pt,bbury=745pt,clip=}}
}
\caption{\bf Field  of GRB970508 optical counterpart from the 6-m telescope.
}
\label{field}
\end{figure*}

\begin{table*}[t]
\begin{center}
\caption{\bf  Coordinates and magnitudes of
secondary standard stars.}
\begin{tabular}{ccccccc}
\hline
 NN &   $\alpha_{2000.0}$&  $\delta_{2000.00}$ &   $\bf B$   &  $\bf V$        & $\bf R_c$       & $\bf I_c$             \\  \hline
  1 &  06:53:37.19     & 79:17:30.7            &   20.44  &   19.14 &  18.31 &  17.53   \\
  2 &  06:53:36.30     & 79:15:30.0            &   19.93  &   19.17 &  18.71 &  18.27   \\
  3 &  06:53:39.23     & 79:15:21.1            &   17.94  &   17.40 &  17.06 &  16.71   \\
  4 &  06:53:48.50     & 79:16:32.7            &   21.93  &   20.43 &  19.49 &  18.53   \\ \hline
\end{tabular}
\end{center}
\label{Filt}
\end{table*}

\clearpage
\begin{figure*}[t]
   \vbox{\psfig{figure=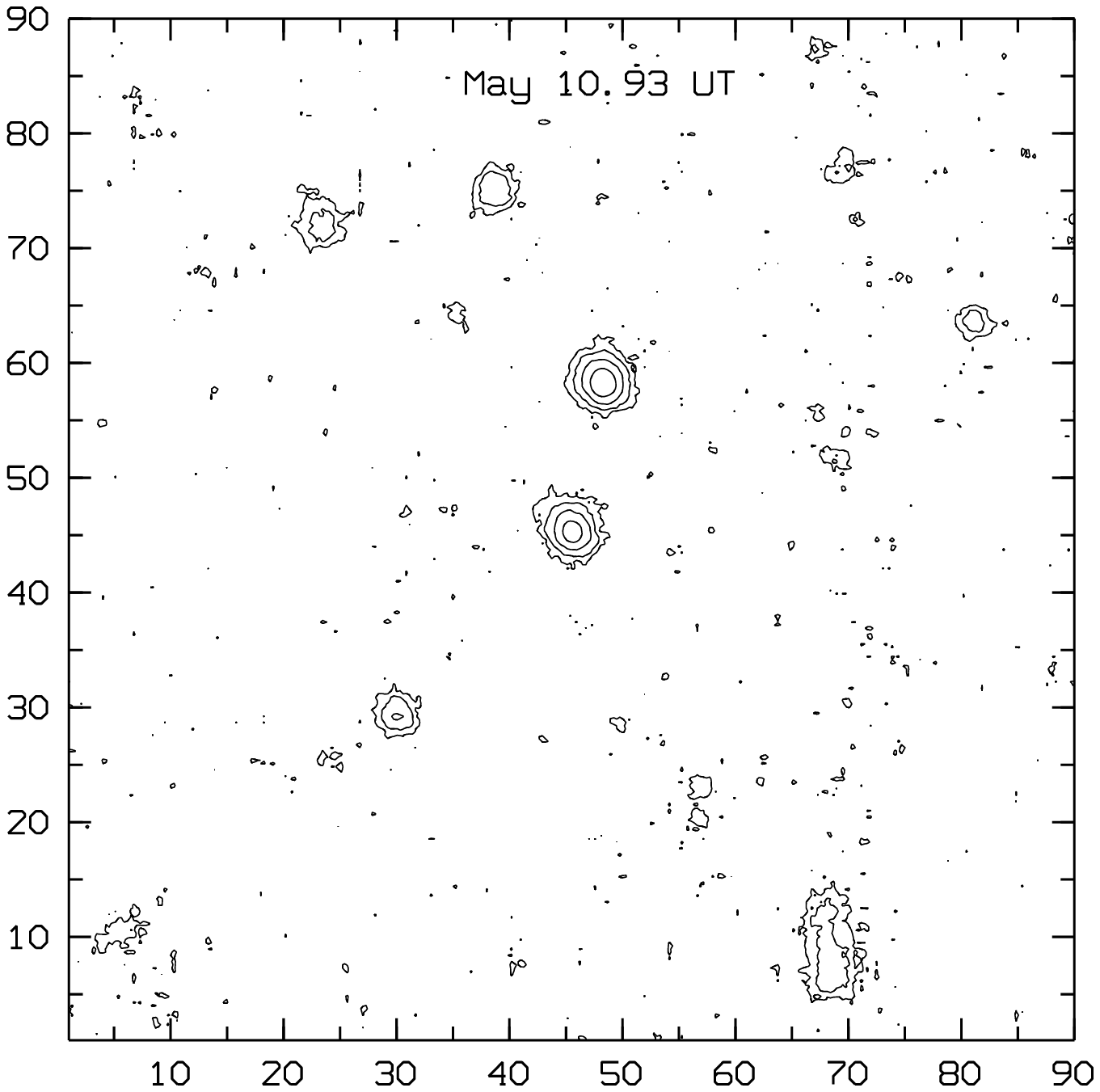,width=9cm,%
 bbllx=50pt,bblly=60pt,bburx=485pt,bbury=485pt,clip=}} \par
\vspace*{-8.7cm}\hspace*{9cm}
  \vbox{\psfig{figure=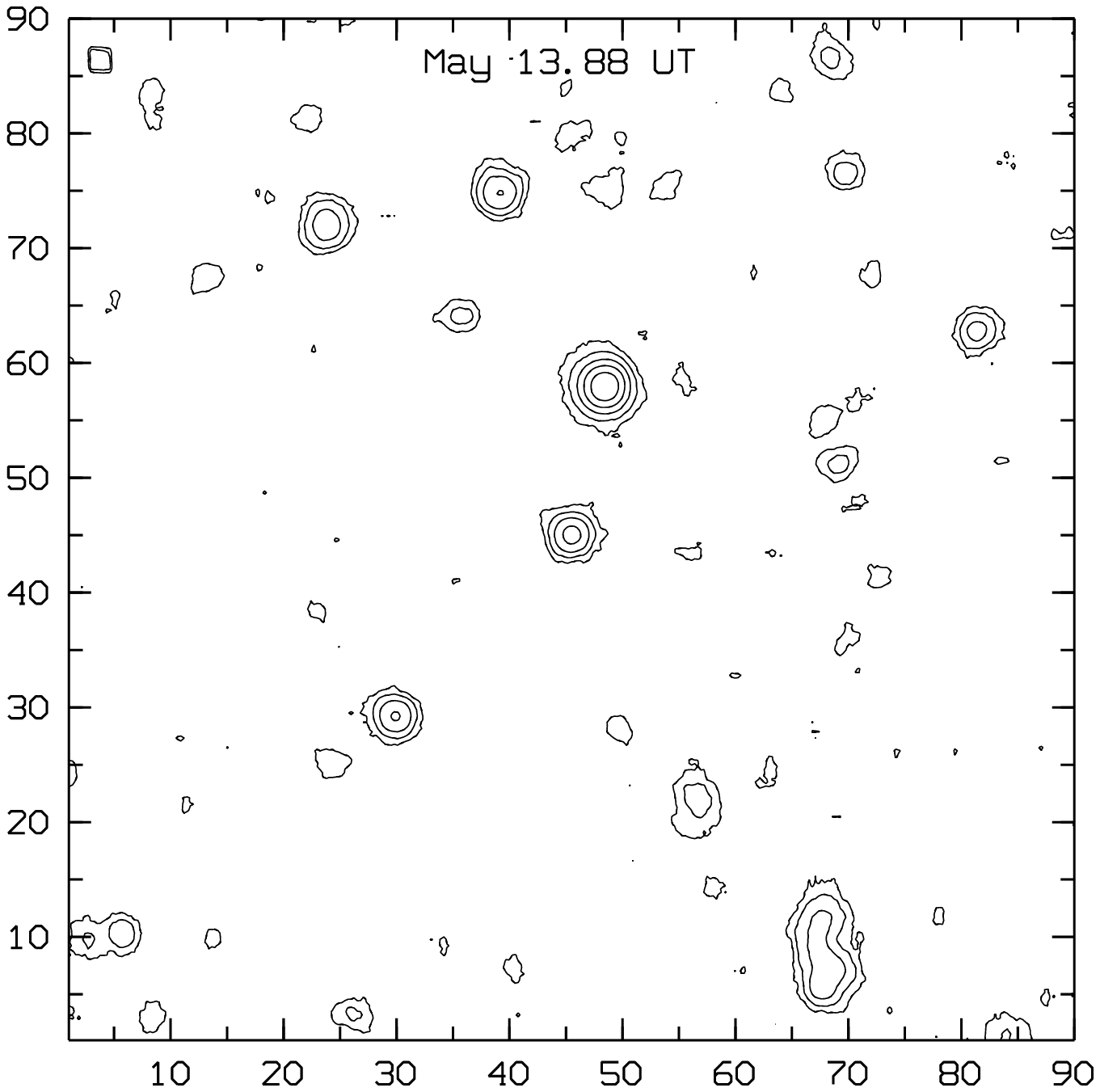,width=9cm,%
 bbllx=50pt,bblly=60pt,bburx=485pt,bbury=485pt,clip=}} \par
   \vbox{\psfig{figure=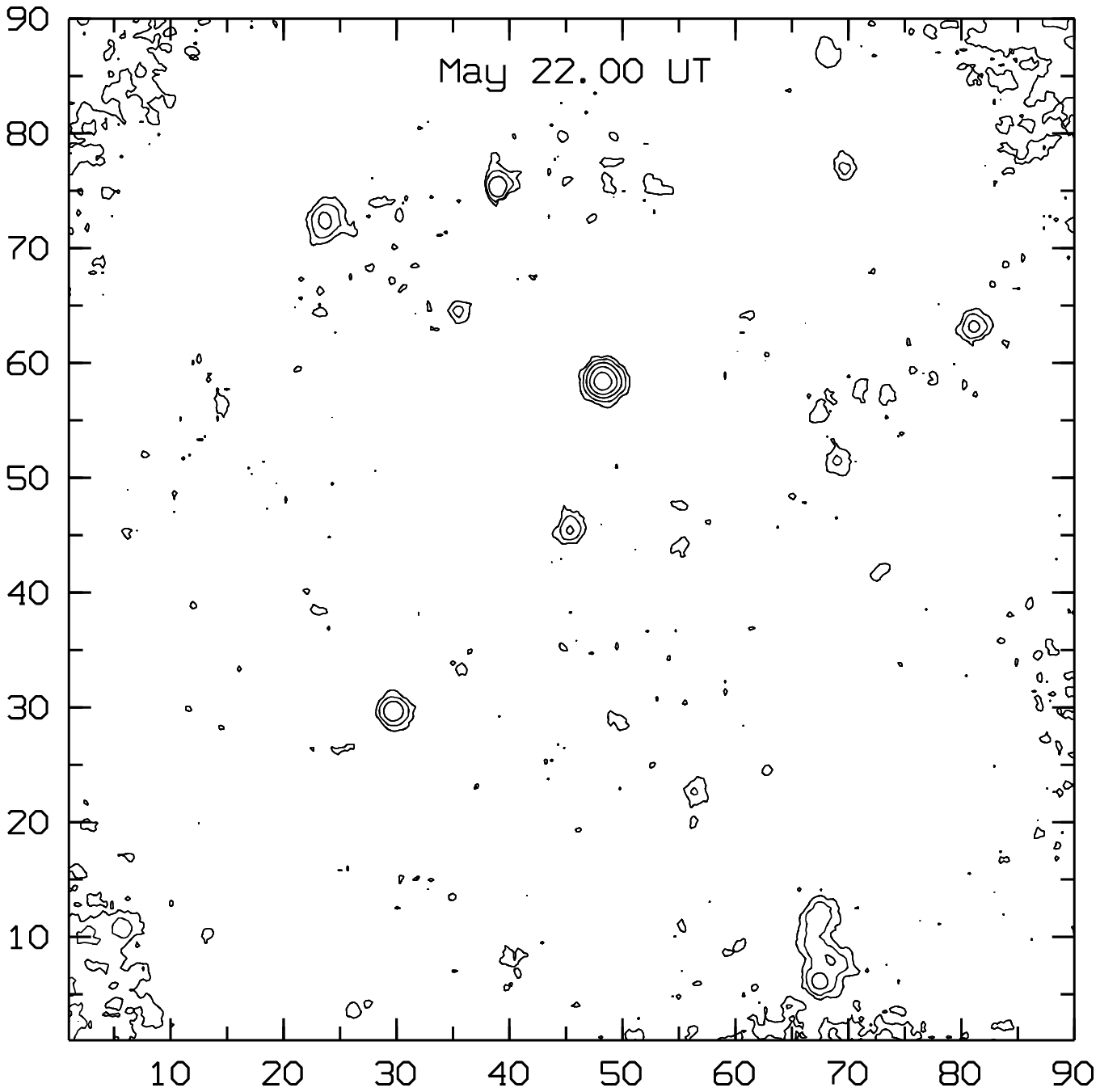,width=9cm,%
  bbllx=50pt,bblly=60pt,bburx=485pt,bbury=485pt,clip=}} \par
\vspace*{-8.7cm}\hspace*{9cm}
   \vbox{\psfig{figure=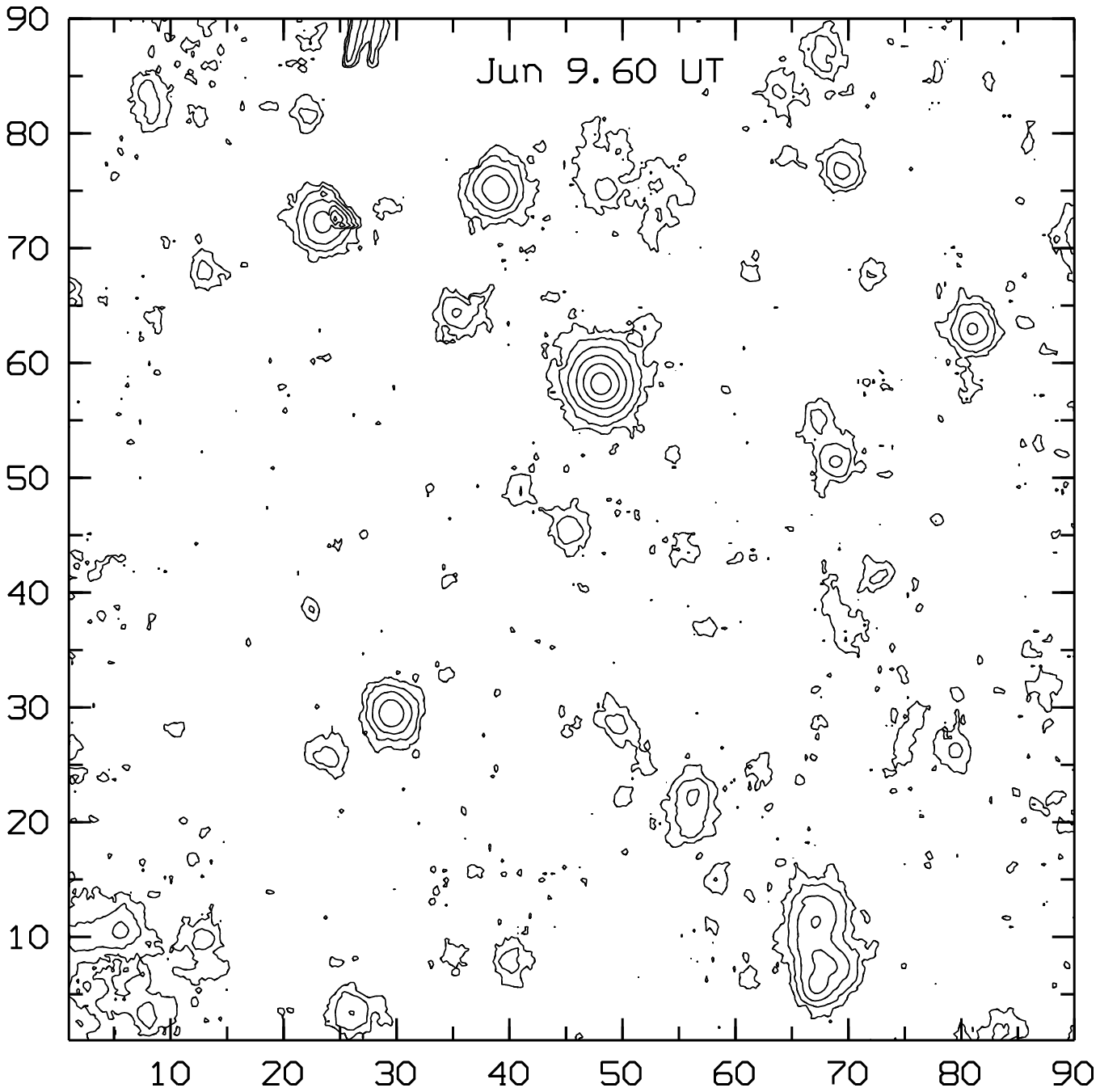,width=9cm,%
  bbllx=50pt,bblly=60pt,bburx=485pt,bbury=485pt,clip=}} \par
\caption{\bf 
Contour plots of the optical counterpart vicinity in $\bf R_c$
obtained  on May 10.93 UT (100s), May 13.88 (450s), May 22.00 (1300s) and
June 09.60 (4000s).
The size of each fragment is $90^{\prime\prime}\times90^{\prime\prime}$.
Optical counterpart of GRB970508 is in the center of each frame.
The lowest level corresponds to 0.75$ F_{sky}^{\frac{1}{2}}$ , where
$F_{sky}$ is a sky flux per one pixel. Next contours are factors 2.5
apart.
}
\end{figure*}

\clearpage
\begin{figure*}[t]
\vspace{13.0cm}
\includegraphics{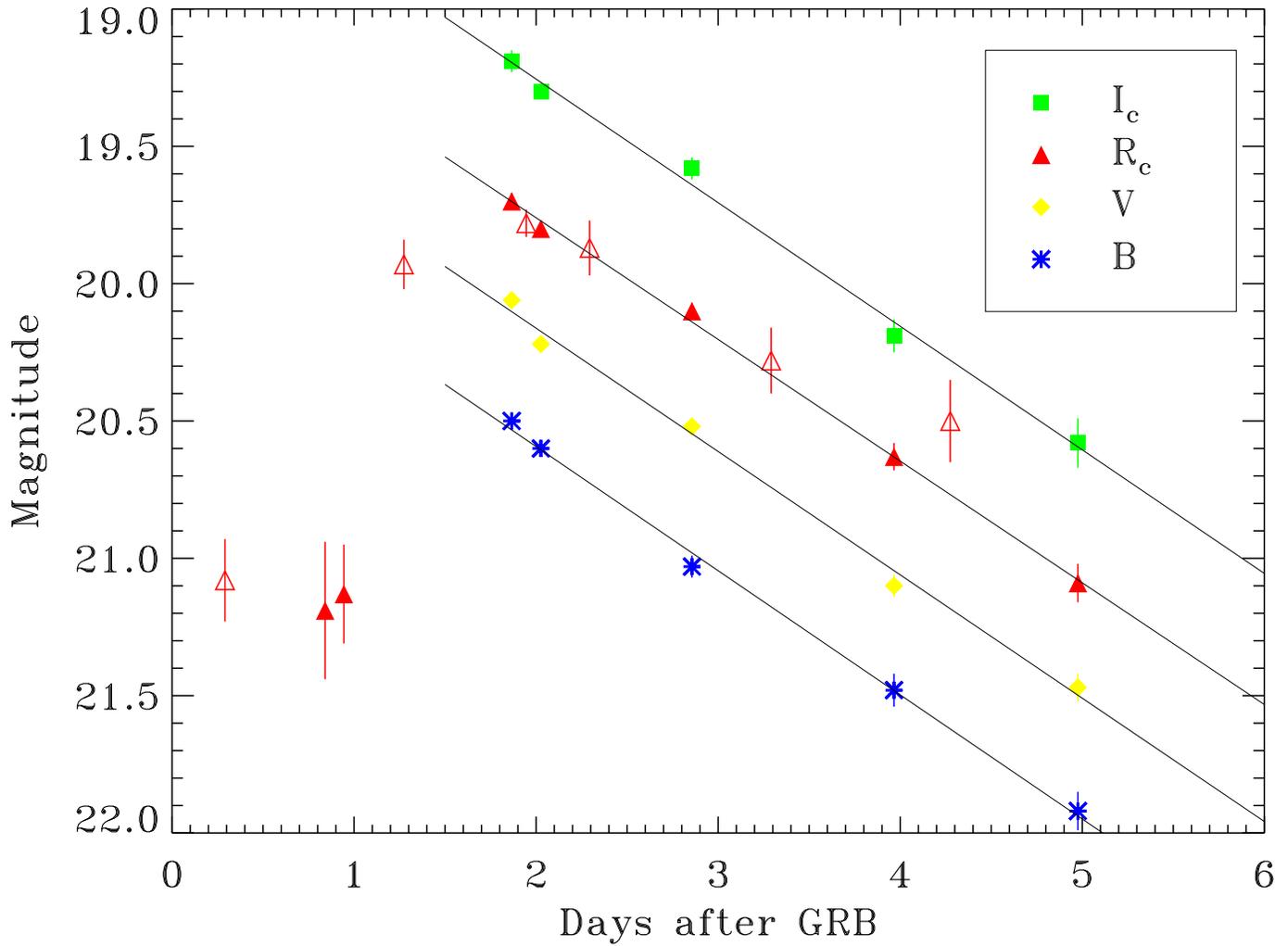}
\caption{\bf  The light curves of GRB970508 optical counterpart during
5 days after the burst.
SAO RAS (filled symbols),
and Loiano (Mignoli M. et al., 1997) ($\bf t-t_0 = 1.95$) and Palomar
(Djorgovski et al., 1997) (transformed to $\bf R_{c} = r - 0.34 + A_{r}$)
(open triangles) magnitudes with their errors are shown.
Lines correspond to the equations of exponential decline of brightness
reported in the text.
}
\label{5day}
\end{figure*}

\begin{figure*}[t]
\vspace{13.0cm}
\includegraphics{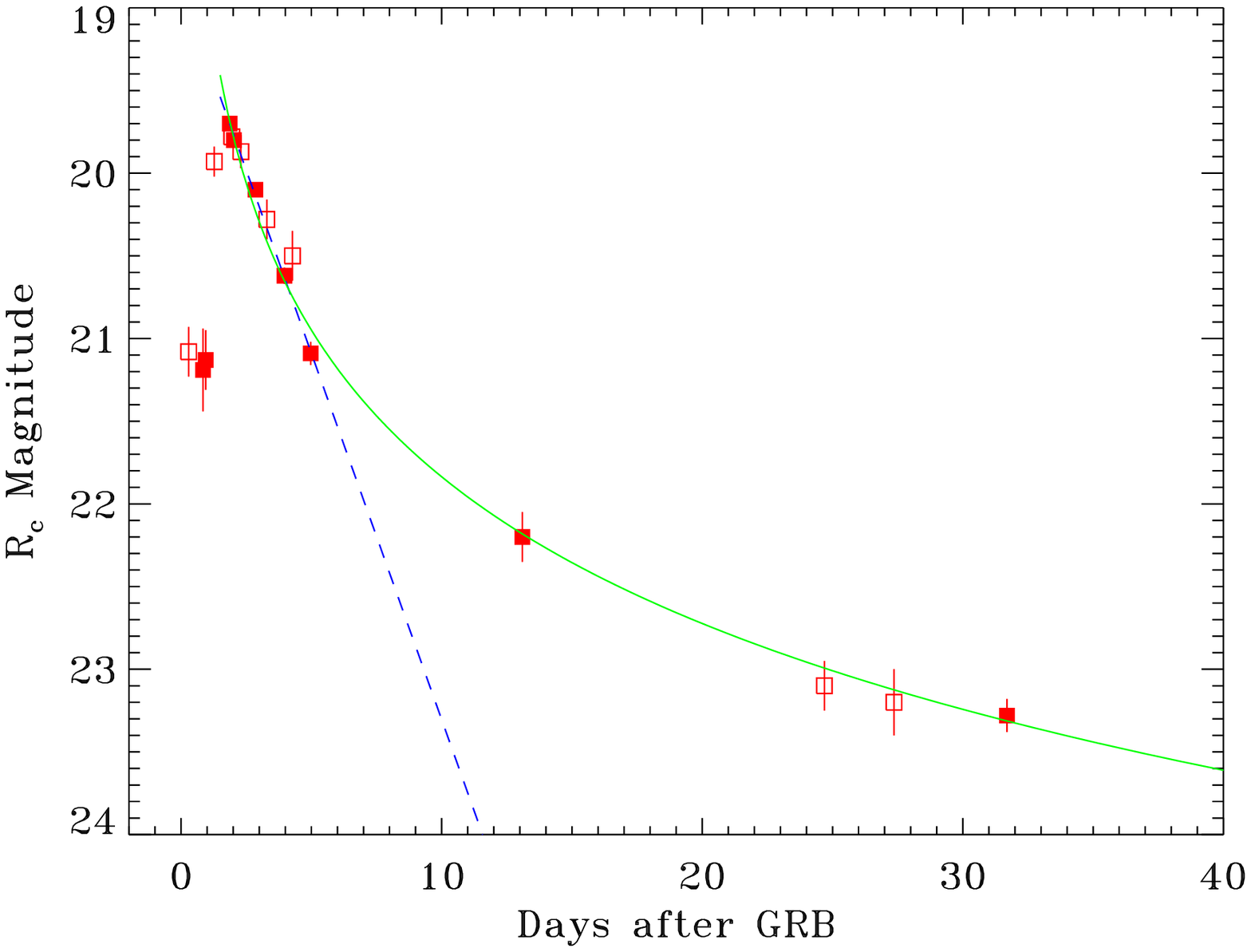}
\caption{\bf $\bf R_c$ light curve of the optical counterpart of GRB970508
during 40 days after the burst.
SAO RAS (filled squares) and 
Palomar (Djorgovski et al., 1997) (transformed to
 $\bf R_{c} = r - 0.34 + A_{r}$),
 HST (Fruchter et al., 1997),
Keck II (Metzger et al., 1997)  (open squares) 
(transformed from Schaefer's photometric system to ours)
 magnitudes are shown.
Lines corresponds to exponential law and power law for fading brightness.
}
\label{fullc}
\end{figure*}
\clearpage
\pagestyle{empty}



\begin{center}
 {\LARGE \bf { UP-DATE!}} \\
\Large \bf { $\bf R_c$ light curve of \underline{GRB970508} optical remnant
 up to 85 day after the burst}
\end{center}
\vspace{-1cm}

\begin{figure}[h]
\vspace{13.0cm}
\includegraphics{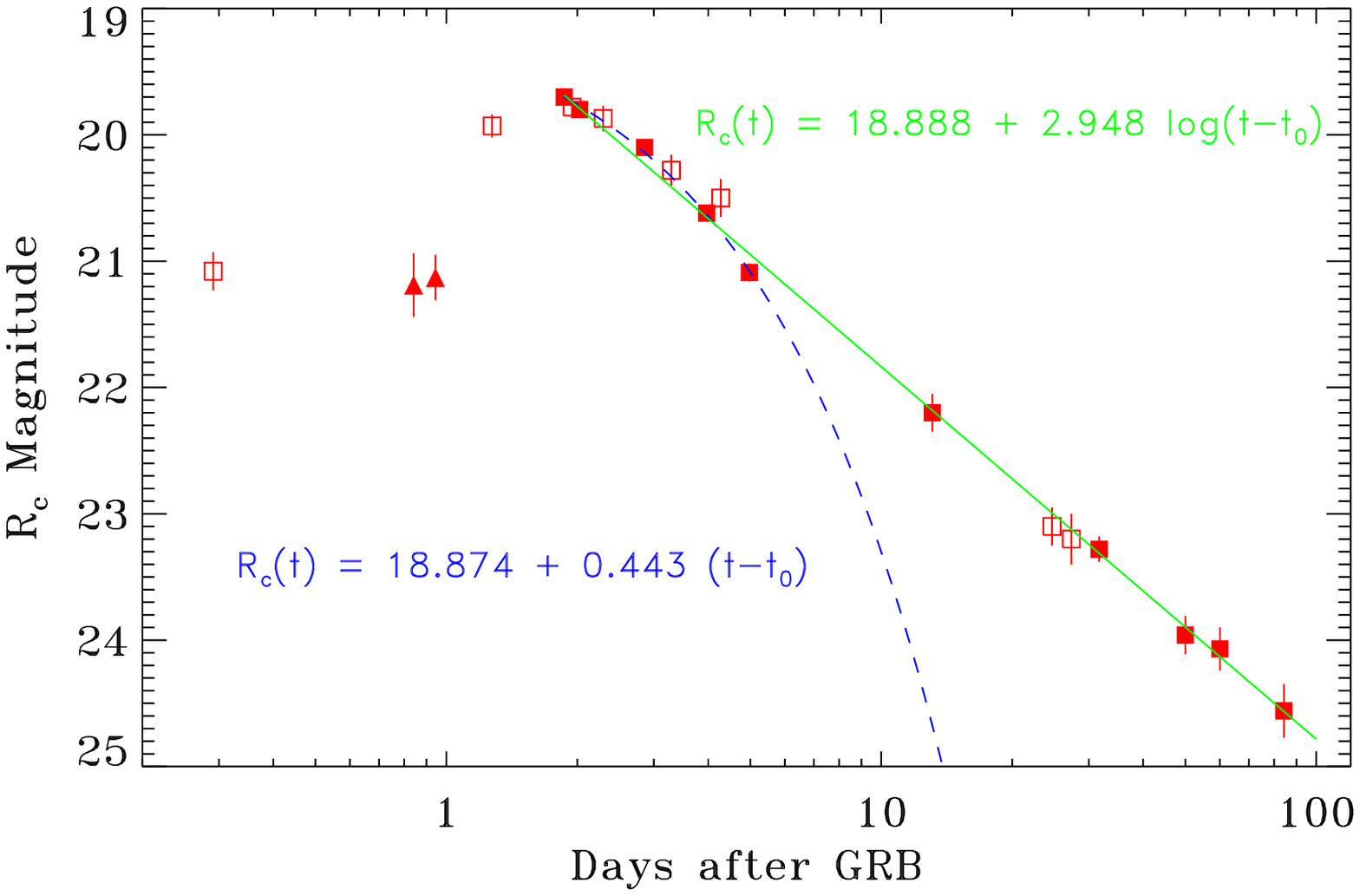}
\end{figure}
\vspace{-1cm}

{\normalsize \bf SAO RAS (including the new data obtained in June,
 July and August, filled squares), Palomar (Djorgovski et al., 1997), 
Loiano (Mignoli et al., 1997), HST (Fruchter et al., 1997),
Keck II (Metzger et al., 1997) (open squares)
(transformed from Schaefer's photometric system to ours)
magnitudes are shown.
Dashed line (using only 6-m telescope data) corresponds to exponential law
for fading brightness (1) up to $\bf  t-t_o=4.976$.
Solid line (using only 6-m telescope data) corresponds to power law for fading brightness (2) up to $\bf t-t_o=84.246$.
}

 \bf
1) Exponential law:
    $$\large \bf  R_c = 18.874(\pm0.029)+0.443(\pm0.011) \;(t-t_o) $$
2) Power law:
  $$\large \bf  R_c = 18.888(\pm0.078)+2.948(\pm0.040) \;log(t-t_o) $$
\medskip
   \centerline{\fbox{{ $\Large \bf  \alpha = 1.179 \;(\pm 0.016) $}}}


\end{document}